\documentstyle[prl,epsf,floats,aps,amsfonts,twocolumn]{revtex}

\input{epsf.tex}
\def\inseps#1#2{\def\epsfsize##1##2{#2##1} \centerline{\epsfbox{#1}}}

\newcommand {\be}{\begin{equation}}
\newcommand {\ee} {\end{equation}}
\newcommand {\ba}{\begin{eqnarray}}
\newcommand {\ea} {\end{eqnarray}}

\def \F2 {FPL${}^2$ }

\def \OMIT #1{}
\def \rem #1 {{\it #1}}

\begin{document}

\twocolumn[\hsize\textwidth\columnwidth\hsize\csname
@twocolumnfalse\endcsname

\title {Exponents and bounds for uniform spanning trees in $d$ dimensions}
\author {N. Read}

\address{Department of Physics, Yale University, P.O. Box
208120, New Haven, CT 06520-8120, USA}

\date{May 10, 2004}
\maketitle

\begin{abstract}
Uniform spanning trees are a statistical model obtained by taking
the set of all spanning trees on a given graph (such as a portion
of a cubic lattice in $d$ dimensions), with equal probability for
each distinct tree. Some properties of such trees can be obtained
in terms of the Laplacian matrix on the graph, by using Grassmann
integrals. We use this to obtain exact exponents that bound those
for the power-law decay of the probability that $k$ distinct
branches of the tree pass close to each of two distinct points, as
the size of the lattice tends to infinity.

\end{abstract}
]


In graph theory, a tree on a graph is a connected subset of the
vertices and edges without cycles, and a spanning tree is a tree
that includes all $n$ vertices of the graph (it must then have
$n-1$ edges). Results for the number of spanning trees on a given
graph go back to the nineteenth century (see e.g.\ Ref.\
\cite{vlw}). If each spanning tree is given equal probability, we
obtain uniform spanning trees. In this paper, we consider uniform
spanning trees on (a portion of) the square, cubic, or hypercubic
lattice in $d$ dimensions. One would like to characterize the
fractal properties of the trees as the size (number of vertices)
of the lattice goes to infinity. One characteristic is the
probability that two well-separated points are nearly connected by
$k=1$, $2$, $3$, \ldots distinct branches of the tree, or
alternatively by distinct paths along the tree, and these are
expected to behave as power laws that are described by critical
exponents. We will study these by methods based on the classical
results, and obtain some exact exponents, which serve as bounds
for more general ones. (In two dimensions, the exact results have
been known for some time \cite{dup,sal92,Maj,ivash,ken}.) The
motivation to consider this problem came from its connection to
some optimization problems, which are in turn connected with the
ground states of classical systems with quenched disorder, such as
Ising spin glasses. In the two-dimensional case, there is also a
connection with loop models, the $Q\to0$ Potts model, and Coulomb
gases in conformal field theory \cite{dup,sal92,Maj}.

First we note that the result [variously attributed either to
Kirchhoff (1847), or to Sylvester (1857), Borchardt (1860), and
Cayley (1856)] for the number $\cal N$ of spanning trees on a
graph can be written in the following generalized form: %
\be {\cal N} = {\rm cof}\, \Delta^{(x_1,y_1)} =
(-1)^{x_1+y_1}\det \Delta^{(x_1,y_1)}, %
\ee where we have recalled the definition of the cofactor. Here
$x_1$, $y_1=1$, $2$, \ldots label the vertices in the graph, the
matrix $\Delta$ (the
lattice Laplacian) is defined as %
\be %
\Delta(x,y)=\left\{\begin{array}{l} {\rm deg}\,x \hbox{ if $x=y$,}\\
                                    -t\hbox{ if $x$ and $y$ are
                                    connected by $t$ edges,}\\
                                    0\hbox{ otherwise,}\end{array}\right.
                                    \ee
and $\Delta^{(x_1,y_1)}$ means the minor of $(x_1,y_1)$, i.e.\
$\Delta$ with row $x_1$ and column $y_1$ deleted. For $x_1=y_1$,
this reduces to the better known result. The effect of deleting a
row and column is to remove the zero mode that would otherwise
cause the determinant of $\Delta$ to vanish.

The result generalizes further to a relation that involves the
number ${\cal N}^{(x_1y_1,x_2y_2,\ldots,x_ky_k)}$ of spanning
subgraphs without circuits with $k$ components, and with $x_i$,
$y_i$ in the same component for each $i$ (we will assume that all
$x_i$, $y_i$ are distinct). The result is%
\ba %
\lefteqn{{\rm cof}\,\Delta^{(x_1\ldots x_k,y_1\ldots
y_k)}=}&&\nonumber\\&&\quad\pm \sum_{P\in S_k} {\cal
N}^{(x_1y_{P(1)},x_2y_{P(2)},\ldots,x_ky_{P(k)})}{\rm sgn}\,P.
\label{cof}
\ea%
Here the cofactor is again %
\be%
(-1)^{\sum_{i=1}^k(x_i+y_i)}\det \Delta^{(x_1\ldots x_k,y_1\ldots
y_k)},%
\ee where rows $x_i$ and columns $y_i$ have been deleted, and $P$
runs over permutations of $k$ symbols. The overall sign on the
right hand side depends on the details of how the vertices are
labelled and is uninteresting. Both the generalizations are
mentioned by Ivashkevich \cite{ivash} (see also Ref.\
\cite{prie}), but he omits the signs in the cofactors. The results
can be proved by an extension of the proof given for example in
Ref.\ \cite{vlw}.

In the following we will consider a graph that is a bounded
portion $\Lambda$ of the $d$-dimensional cubic lattice ${\bf Z}^d$
(with edges that connect only nearest neighbors at Euclidean
distance 1 in lattice units). We will be interested in the
following property of a spanning tree. We choose two vertices $x$,
$y$, together with a neighborhood of each. We assume that the
neighborhoods are chosen in such a way that the boundary passes
through some vertices, but no edges of $\Lambda$ cross the
boundary; all edges are either inside or outside. We take $k$
vertices $x_i$ on the boundary of the neighborhood $x$, and $k$
vertices $y_i$ on the boundary of that of $y$.  In practise, this
can be satisfied using neighborhoods that are approximately balls
of radius of order $k^{1/(d-1)}$. We can now look at the part of
the tree lying outside the two neighborhoods; this amounts to a
forest of trees, with each tree rooted on both the boundaries of
the neighborhoods of $x$ and $y$. We ask whether, for each $i$,
the points $x_i$, $y_i$ lie in the same connected component in
this forest, and are in a distinct component from any other pair
$x_j$, $y_j$. If so, then in terms of the original tree the $x_i$s
are connected to the corresponding $y_i$s by branches of the tree
that are distinct outside the two neighborhoods selected, and we
call this ``crossing $k$ times (between specified points in the
neighborhoods of $x$ and $y$) by distinct branches''. See Fig.\ 1.
For given $\Lambda$, neighborhoods of $x$ and $y$, and points
$x_i$, $y_i$, denote the number of such trees ${\cal
N}^{(x_1y_1,x_2y_2,\ldots,x_ky_k)}_{\rm branches}$. Then we also
define corresponding probabilities,
\be {\cal P}^{(x_1y_1,x_2y_2,\ldots,x_ky_k)}_{\rm branches}={\cal
N}^{(x_1y_1,x_2y_2,\ldots,x_ky_k)}_{\rm branches}/{\cal N}.\ee

\begin{figure}
\inseps{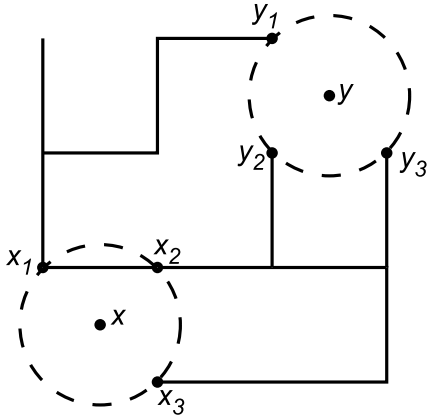}{0.6} \caption{Example of a spanning tree on a
portion of the square lattice, with points $x$, $y$ and a
neighborhood of each marked, with vertices $x_1$, $x_2$, $x_3$ and
$y_1$, $y_2$, $y_3$ on the boundary of each shown. In this
example, there are paths, lying outside the neighborhoods, that
connect $x_i$ to $y_i$ for each $i$, as required, and those paths
are distinct, but there are only two distinct branches (connected
components) outside the neighborhoods. The pairs $x_2$, $y_2$ and
$x_3$, $y_3$ do not lie on distinct branches.} \label{}
\end{figure}

It is obvious that there is a relation between this definition,
and the $k$-component spanning subgraphs without circuits
considered before, if the latter are defined on $\Lambda^-$, that
is $\Lambda$ with the interiors of the neighborhoods of $x$ and
$y$ removed. Given a $k$-component spanning subgraph of
$\Lambda^-$, a spanning tree of $\Lambda$ can be obtained by
adding $k-1$ edges, each inside either the neighborhood of $x$ or
of $y$, together with one edge for each vertex in
$\Lambda-\Lambda^-$. If we also assume that there are exactly $k$
vertices on the boundaries of each of the two neighborhoods, then
${\cal N}^{(x_1y_1,x_2y_2,\ldots,x_ky_k)}_{\rm branches}$ is
obtained by summing over all the ways to add edges to each
$k$-component spanning subgraph of $\Lambda^-$ (with $x_i$ and
$y_i$ in the $i$th component for all $i$) to obtain a spanning
tree of $\Lambda$ whose branches are distinct outside the two
neighborhoods, and then also summing over the $k$-component
subgraphs used in this construction. As the maximum possible
number of ways to add edges is limited by the size of the
neighborhoods used, ${\cal N}^{(x_1y_1,x_2y_2,\ldots,x_ky_k)}_{\rm
branches}$ is only slightly larger than the number ${\cal
N}^{(x_1y_1,x_2y_2,\ldots,x_ky_k)}$ of $k$-component spanning
graphs of $\Lambda^-$. (More precisely, there is a bound ${\cal
N}^{(x_1y_1,x_2y_2,\ldots,x_ky_k)}_{\rm branches}\leq c_{k,d}
{\cal N}^{(x_1y_1,x_2y_2,\ldots,x_ky_k)}$ where $c_{k,d}$ is
independent of the distance from $x$ to $y$.)

Another possible definition of crossing from $x$ to $y$ would
require only that the $k$ crossings of the spanning tree be made
by $k$ {\em paths} on the tree that are distinct (have no edges in
common) outside the neighborhoods (for $k=1$ these are the same
thing, and the probability is one). See Fig. 1. These numbers
${\cal N}^{(x_1y_1,x_2y_2,\ldots,x_ky_k)}_{\rm paths}$ can be
obtained by adding edges to the $k$-component spanning subgraph of
$\Lambda^-$ in arbitrary positions in $\Lambda$, and so ${\cal
N}^{(x_1y_1,x_2y_2,\ldots,x_ky_k)}_{\rm branches}\leq{\cal
N}^{(x_1y_1,x_2y_2,\ldots,x_ky_k)}_{\rm paths}$, and
\ba%
{\cal N}^{(x_1y_1,x_2y_2,\ldots,x_ky_k)}/{\cal N}&\leq& {\cal
P}^{(x_1y_1,x_2y_2,\ldots,x_ky_k)}_{\rm
branches}\nonumber\\&\leq&{\cal
P}^{(x_1y_1,x_2y_2,\ldots,x_ky_k)}_{\rm paths}.%
\ea%
These continue to hold even if there are more than $k$ vertices on
the boundary of each neighborhood.

We will be interested in the scaling limit in which we first let
$\Lambda\to {\bf Z}^d$ (i.e.\ the system size tends to infinity)
with $x$, $y$ fixed, then let the Euclidean distance $|x-y|$ (in
lattice units) become large. In this limit, the leading behavior
of any of the ratios or probabilities just defined will behave as
a power law, ${\cal P}\propto |x-y|^{-2X_k}$, where $X_k$ is a
scaling dimension (which also depends on $d$). However, in the
antisymmetrized combination that appears in eq.\ (\ref{cof}), the
leading part of
\be%
{\cal N}^{(x_1y_1,x_2y_2,\ldots,x_ky_k)}/{\cal N}\propto
|x-y|^{-2X_k^{k\,{\rm components}}}\ee%
may be cancelled, so that a subleading power (with a larger
exponent $X_k^{\rm antisymm}$) may be dominant. Thus in
general the exponents must obey the inequalities %
\be%
X_k^{\rm antisymm}\geq X_k^{k\,{\rm components}}\geq X_k^{\rm
branches}\geq X_k^{\rm paths}.\label{ineq}%
\ee%

As the addition of edges in the neighborhood of $x$ or $y$
contributes only a constant factor to the numbers ${\cal N}_{\rm
branches}$ , we have that $X_k^{k\,{\rm components}}= X_k^{\rm
branches}$. For the two-dimensional case, the tree branches must
enter the neighborhood of $x$ in a sequence, and similarly at $y$.
As the tree branches that cross from $x$ to $y$ cannot intersect
because $\Lambda$ is planar, the only nonzero terms in eq.\
(\ref{cof}) are for $P$s that differ only by a cyclic permutation.
For $k$ odd, all cyclic permutations of $k$ objects are even
permutations, so all terms have the same sign, as if the sum were
symmetrized. Therefore in $d=2$, $X_k^{\rm antisymm}=X_k^{k\,{\rm
components}}$ for $k$ odd. We also note that in $d=2$ dimensions,
the boundary of a ``thickened'' tree is a nonintersecting dense
loop, and the crossing by distinct branches corresponds to
crossing by the loop, $2k$ times.

There is one further subtlety that must be mentioned. We have
defined ratios and probabilities on finite graphs, followed by an
infinite-volume limit. Each finite graph is spanned by a single
tree, by construction, but as the volume increases, a path from
one vertex to another in a typical tree may involve larger and
larger excursions, so that for infinite volume, two vertices might
be connected only ``at infinity'', and then they can be regarded
as not connected. Then the limiting measure is for a forest of
trees, not a single tree. It turns out that for $d\leq 4$, there
is a single tree in the infinite-volume limit, while for $d>4$
there are infinitely many (infinite-size) trees \cite{pem} (these
statements hold with probability one). Crossing probabilities by
$k$ distinct branches or $k$ distinct paths can be defined here as
well, with crossing not allowed to be ``through infinity''; we
denote exponents under this condition by ``$<\infty$''. These
probabilities will be less than or equal to the corresponding ones
defined above, and so the related exponents obey $X_k^{{\rm
branches}<\infty}\geq X_k^{\rm branches}$, $X_k^{{\rm
paths}<\infty}\geq X_k^{\rm paths}$. One has
$X_1^{<\infty}=(d-4)/2$ for $d>4$ (Ref.\ \cite{pem}, Thm.\ 4.2),
which because of the way this probability is normalized (any
vertex is on some tree) is consistent with the belief that the
Hausdorff dimension of any of the trees in the forest is $4$ for
$d>4$. It would be interesting to obtain the remaining exponents
$X_k^{{\rm paths}<\infty}$ and $X_k^{{\rm branches}<\infty}$ for
$k>1$ also. In any case, the distinction in definition disappears
for $d\leq 4 $ where there is a single spanning tree.

We now turn to the calculation of the antisymmetrized combinations
of $\cal N$s. According to eq.\ (\ref{cof}), the antisymmetrized
sum of the relevant ratios is given by a ratio of cofactors:%
\ba %
\lefteqn{\frac{{\rm cof}\,\Delta^{(x_1\ldots x_k,y_1\ldots
y_k)}}{{\rm cof}\,\Delta^{(x_1,y_1)}}}\nonumber\\&&\quad=\pm
\sum_{P\in S_k} {\cal
N}^{(x_1y_{P(1)},x_2y_{P(2)},\ldots,x_ky_{P(k)})}
{\rm sgn}\,P/{\cal N}.%
\ea%
As is well-known, such cofactors can be obtained from Gaussian
integrals over Grassmann variables $\psi$, $\psi^\ast$,%
\ba%
\lefteqn{{\rm cof}\,\Delta^{(x_1\ldots x_k,y_1\ldots
y_k)}}&&\nonumber\\\quad&=&\pm\int\prod_xd\psi_xd\psi^\ast_x\,
\psi_{x_1}\cdots\psi^\ast_{y_k}
e^{\sum_{x,y}\psi_x\Delta(x,y)\psi^\ast_y},%
\ea%
where the overall sign is determined by the order of the Grassmann
integrations (not by the selected values of $x_i$, $y_i$). In the
limits $\Lambda\to {\bf Z}^d$, followed by the limit of $x$ and
$y$ far apart, the cofactors above become Gaussian integrals for a
continuum massless complex scalar Fermi field. The equation of
motion for the Fermi field $\psi$ is simply $\Delta \Psi =0$,
where $\Delta =\sum_{\mu=1}^d\partial_\mu\partial_\mu$ is the
Laplacian in $d$ dimensions. As all $x_i\to x$, the ratio of
cofactors becomes a sum of
correlation functions of operators of the form%
\be {\cal O}(x)=\psi\partial_\mu\psi\cdots\partial_{\mu_1\mu_2\cdots}\psi%
\ee%
(with $k$ $\psi$s, and where
$\partial_{\mu_1\mu_2\cdots}=\partial_{\mu_1}\partial_{\mu_2}\cdots$)
at $x$, with a similar operator at $y$ with $\psi^\ast$ in place
of $\psi$. The undifferentiated $\psi$ and $\psi^\ast$ are
necessary to cancel the zero mode, both in the numerator and
denominator of the correlation function. The remaining integrals
can be simply expressed (using Wick's theorem) in terms of sums of
products of derivatives of Green's functions
$G(x,y)=\Delta^{-1}(x,y)$ for the scalar field $\psi$ in $d$
dimensions, and the required scaling limit of this expression
exists; one has $G(x,y)\propto |x-y|^{d-2}$ for $d>2$. Thus the
scaling dimension of $\psi$ or $\psi^\ast$ is $(d-2)/2$ in $d$
dimensions, and an operator of the above form $\cal O$ has scaling
dimension $X_k^{\rm antisymm}={\rm dim}\,{\cal O}$ equal to
$(k-1)(d-2)/2$ plus the number of partial derivatives in $\cal O$
(note that we replaced $k$ by $k-1$ because the subtracted zero
mode does not contribute to scaling). This implies that for $k=1$,
the operator has dimension zero, which is correct as a spanning
tree connects any two points $x$, $y$. It will be convenient to
define ${\rm dim}'\,{\cal O}={\rm dim}\,{\cal O}-(k-1)(d-2)/2$.

To find the operator that contributes the leading behavior of the
correlation function for a given $k$, we must use as few
derivatives as possible. Further, the equation of motion implies
that any trace such as $\sum_\mu\partial_{\mu\mu\mu_3\cdots}\psi$
vanishes. Then the leading term $\cal O$ is a product in which
each multi-index partial derivative
$\partial_{\mu_1\mu_2\ldots}\psi$ is a traceless symmetric tensor,
and the total degree (number of derivatives) in the product is as
low as possible. Because of the anticommutation of the $\psi$s,
$\cal O$ vanishes unless the traceless symmetric tensors
$\partial_{\mu_1\mu_2\ldots}\psi$ are linearly independent. We
notice that the traceless symmetric tensors of given rank (degree)
in dimension $d$ form an irreducible representation of the
rotation group in $d$ dimensions, SO($d$). In general, we expect
that the leading part of the crossing probability ${\cal P}_k$
transforms as a scalar under rotations. A scalar
(rotationally-invariant) operator can be obtained if the product
includes either all or none of the members of a complete
linearly-independent set of traceless symmetric tensors for each
degree (rank) $l$. Making the minimal choices of the degrees, we
notice that this is analogous to filling states for fermions,
where the single fermion states correspond to the symmetric
traceless tensors. These tensors transform the same way as
``hyperspherical harmonics'', which span the space of functions on
a sphere $S^{d-1}$. This can be seen easily by representing each
$\partial_\mu$ by a component of a vector $x_\mu$, which
transforms the same way, and the traces can be excluded if we
assume that $\sum_\mu x_\mu x_\mu$ is constant, so that tensors
with non-vanishing trace are equivalent to lower-degree functions.
Then the symmetric functions in $x_\mu$ under this condition are
simply the functions on $S^{d-1}$. If each fermion on $S^{d-1}$
has a kinetic energy that is equal to the angular momentum $l$,
then the many-fermion state with lowest total kinetic energy for
given number of fermions $k$ corresponds to using the lowest total
degree. The case where all traceless symmetric tensors of each
(lowest) degree are used corresponds to filling a Fermi sea by
filling the lowest shells up to angular momentum (degree of the
traceless symmetric tensor) $L$. The total kinetic energy
corresponds to the scaling dimension ${\rm dim}'\,{\cal O}$. When
the lowest states are all filled, but the topmost shell is only
partially filled, the scaling dimension interpolates linearly
between the values it takes for filled shells. We point out that
the use of fermions on the sphere is more than an analogy, as the
field theory of fermions on $S^{d-1}$ with time $t$ corresponds to
the original problem in radial quantization, obtained by
conformallly mapping ${\bf R}^d$ to $S^{d-1}\times {\bf R}$ by a
logarithmic change of variable, so that the dilatation operator
becomes the Hamiltonian for the radial evolution.

We define $N(l,d)$ to be the dimension of the space of traceless
symmetric tensors of degree $l$ in dimension $d$. When the shells
are filled up to degree $L$, the preceding considerations lead
immediately to
the relations%
\ba k&=&\sum_{l=0}^L N(l,d),\label{k}\\
{\rm dim}'\,{\cal O}_L&=&\sum_{l=0}^L l N(l,d).\label{dim'}%
\ea $N(l,d)$ is given for all $l\geq0$ by%
\be N(l,d) = \left(\begin{array}{c}l+d-1\\
                                   l\end{array}\right)
              -\left(\begin{array}{c}l+d-3\\
                                   l-2\end{array}\right).
\ee%
Here the first term is the number of symmetric tensors, and the
subtraction is for removing the traces. From the binomial
coefficients one sees that $N(l,d)$ is a polynomial in $l$ of
degree $d-2$ for $d\geq 2$, and hence $k$ is a polynomial in $L$
of degree $d-1$, and ${\rm dim}'\,{\cal O}_L$ is a polynomial of
degree $d$.

The leading behavior for $l$ large is %
\be%
N(l,d)\sim \frac{2 l^{d-2}}{(d-2)!}%
\ee%
(throughout this paper, we use notation $X\sim Y$ as $Z\to\infty$
in the strict sense: $\lim_{Z\to \infty}X/Y = 1$).
Then we find%
\ba%
k&\sim& \frac{2 L^{d-1}}{(d-1)!},\\
{\rm dim}'\,{\cal O}_L&\sim&\frac{2L^d}{d(d-2)!}\,,%
\ea and hence %
\be%
{\rm dim}'\,{\cal O}_L\sim{\rm dim}\,{\cal O}_L\sim
\frac{d-1}{d}\left[\frac{(d-1)!}{2}\right]^{\frac{1}{d-1}}
k^{\frac{d}{d-1}},\label{xkgrowth}%
\ee%
for the values of $k$ specified. As mentioned above, for other
values of $k$, the scaling dimension (now for an operator with
nonzero spin in general) lies on a piecewise linear continuous
curve that interpolates the values above, and lies above that
given implicitly by eqs.\ (\ref{k}), (\ref{dim'}) as polynomials
in $L$, which are trivially extended to continuous values.

By the inequalities (\ref{ineq}), this result gives only an upper
bound on the leading exponents $X_k^{\rm branches}$ or $X_k^{\rm
paths}$. However, the general formulas do give the exact exponents
$X_k^{\rm antisymm}$ for some, possibly subleading, terms in the
probability. The rate of growth of the dimensions $X_k^{\rm
paths}$ on the tree to cross from $x$ to $y$ was shown rigorously
to be less than of order $k^{d/(d-1)}$ in Ref.\ \cite{ABNW}. We
obtain a bound with the same power, but now with a precise
coefficient, and with subleading corrections. Note that the
piecewise-linear curve for $X_k^{\rm antisymm}$ is very close to
its lower envelope, close enough that they have the same average
rate of growth, eq.\ (\ref{xkgrowth}).

We now consider the exact form of the dimensions obtained here for
the $k$ values given by eq.\ (\ref{k}) in small $d$. In two
dimensions, $N(l,2)=2$
($l>0$), and then %
\be%
{\rm dim}\,{\cal O}_k = (k^2-1)/4%
\ee%
for $k$ odd. This is in agreement with earlier results
\cite{dup,sal92,ivash,ken}. (Note that the scaling exponent for
the path connecting $x$ to $y$ along the tree \cite{Maj}
corresponds to the case $k=2$, by considering the dual tree.) As
emphasized above, for $d=2$ and $k$ odd, the arguments in this
paper produce the {\em exact} $X_k^{\rm antisymm}=X_k^{\rm
branches}=(k^2-1)/4$, not only a bound. Note that after replacing
$k$ by $k/2$, this result is the same as the ``$k$-leg'' dimension
for $k$ crossings by a dense polymer \cite{DupSal87}.

For $d=3$, we have the familiar formula $N(l,3)=2l+1$, and then
$k=(L+1)^2$, so $L=\sqrt{k}-1$. For the scaling dimensions, ${\rm
dim}'\,{\cal O}_L= L(L+1)(4L+5)/6$, and%
\be%
X_k^{\rm antisymm}={\rm dim}\,{\cal O}_k =
\frac{2}{3}k^{3/2}-\frac{1}{6}k^{1/2}-\frac{1}{2}.%
\ee%

For $d>3$, one can similarly solve explicitly, but the results are
not as simple (in particular, they are not polynomials in
$k^{\frac{1}{d-1}}$). I am grateful to C. Tanguy for pointing out
that eqs.\ (\ref{k}), (\ref{dim'}) can be summed in closed form
for all $d$, and that for $d$ odd, $k$ is a polynomial in
$[L+(d-1)/2]^2$ of degree $(d-1)/2$. Hence in the cases $d=4$,
$5$, $7$, and $9$, ${\rm dim}\,{\cal O}_k$ can be expressed in
terms of radicals in $k$. For $d=6$, $8$ and all $d\geq 10$, one
meets polynomials of degree greater than four, and the results
presumably cannot be expressed in terms of radicals.

It is tempting to believe that the results obtained here for the
``filled shell'' values of $k$, and their smooth (polynomial in
$L$) extension to general $k$, might be the exact values of
$X_k^{\rm branches}$ and $X_k^{\rm paths}$ for general dimension
$d>2$, as well as for $d=2$. While it appears quite possible that
$X_k^{\rm branches}=X_k^{\rm paths}$ in general, it is not at all
clear that they equal $X_k^{\rm antisymm}$, especially as the
equality that holds in two dimensions could be obtained from a
simple argument that all cyclic permutations of an odd number of
objects are even, an argument that definitely does not go through
in $d>2$.

In conclusion, we have obtained, essentially rigorously, a precise
upper bound on the exponents for $k$ crossings of the uniform
spanning tree on a finite graph in $d$ dimensions, as well as some
exact scaling dimensions in each dimension. However, we have not
addressed the exponents in the uniform spanning forest which
obtains on an infinite graph for $d>4$.

I thank H. Saleur, J.L. Jacobsen, and C. Tanguy for helpful
correspondence, and S. Sachdev for preparing the figure. This work
was supported by the NSF under grant no.\ DMR-02-42949.



\begin{references}

\bibitem{vlw}         J.H. van Lint and R.M. Wilson, {\it A Course in
                       Combinatorics} (Cambridge University, Cambridge,
                       1992), Chapter 34.

\bibitem{dup}         B. Duplantier, J. Stat. Phys. {\bf 49}, 411
                      (1987); B. Duplantier and F. David, J. Stat. Phys.
                      {\bf 51}, 327 (1988).

\bibitem{sal92}      H. Saleur, Nucl. Phys. B {\bf 382}, 486 (1992).

\bibitem{Maj}        S.N.~Majumdar, \prl {\bf 68}, 2329 (1992).

\bibitem{ivash}       E.V. Ivashkevich, J. Phys. A {\bf 32}, 1691
                       (1999).

\bibitem{ken}         R. Kenyon, J. Math. Phys. {\bf 41}, 1338 (2000).

\bibitem{prie}        V.B. Priezzhev, Sov. Phys. Usp. {\bf 28},
                       1125 (1985).

\bibitem{pem}         R. Pemantle, Ann. Prob. {\bf 19}, 1559
                       (1991).

\bibitem{ABNW}        M.~Aizenman, A.~Burchard, C.~M.~Newman, and
                       D.~B. Wilson, Random Struct. Algorithms. {\bf 15},
                       319 (1999).

\bibitem{DupSal87}    B.~Duplantier and H.~Saleur,
                       Nucl.~Phys.~B {\bf 290}, 291 (1987).



\end{references}
\end{document}